\documentclass[pra,preprint,aps]{revtex4}
\usepackage{graphicx}

\begin{document}

\title{LOW ENERGY (E,2E) IONIZATION OF ARGON IN THE EQUAL ENERGY SHARING
GEOMETRY.}

\author{S. Mazevet\S \\ G. Nguyen Vien \P,
J. Langlois\P, R. J. Tweed \P, O. Robaux \P \\ C. Tannous\dag  \\ 
K. Fakhreddine\ddag}
\affiliation{\S Department of Physics and Astronomy, 440 W. Brooks Norman, OK 73019, USA\\
 \P Laboratoire des Collisions Electroniques et Atomiques, 
Université de Bretagne Occidentale, BP: 809 - 29285 Brest Cedex, France \\
  \dag Laboratoire de Magnétisme de Bretagne, Université de Bretagne Occidentale,
CNRS UPRES A 6135, BP: 809 - 29285 Brest Cedex, France\\
 \ddag Lebanese University and CNRS, BP: 11-8281, Beirut, Lebanon
 }

\begin{abstract}
Quantum Defect theory is a well established theoretical concept in modern spectroscopy.
We show that this approach is useful in electron impact ionization problems where state of the art
theoretical methods are presently restricted mostly to simple atomic targets. For the well 
documented Argon ionization  case in equal energy sharing geometry the  approach suggested  
leads to significant improvements compared to previous calculations.
\end{abstract}

\maketitle

\section{INTRODUCTION}

Quantum Defect (QD) information is widely exploited in modern spectroscopy, to characterize Rydberg states and in the calculation of the photoionization cross sections of various atomic and molecular species \cite{jungen,aymar96}. \\
In the present work, we show that QD information, used within the Distorted Wave Born Approximation (DWBA) framework, might also prove useful for the description of ionization processes by presenting an alternative way to account for the short range interactions (static and exchange) in the calculation of the final state continuum distorted waves. The range of validity of this approach reaches beyond that \cite{riley} of the commonly used Furness-McCarthy local exchange approximation \cite{furness}. Compared to the determination of the Hartree-Fock non-local operator \cite{winkler}, which becomes rapidly a tedious task as the size of the target increases, our method allows for a target-independent procedure which can be readily applied to much larger atomic or molecular systems. \\

The parameters of the Green-Sellin-Zachor \cite{green69} parametric form of the electron-ion potential are optimized in order to reproduce the QD using the canonical function method \cite{kobeissi82}. The canonical function method (CFM) is a powerful means for solving accurately the Radial Schr\"{o}dinger Equation (RSE). The quantum eigenvalues of the RSE are obtained to any desired accuracy without having to evaluate the eigenfunctions. Several parametric potentials are discussed extensively in the literature and we already used several functional forms \cite{aymar96} adapted to different atomic systems. We studied recently \cite{tannous99} the QD of some rare gases with the Klapisch parametric potential \cite{klapisch71} and found that in some cases it was very difficult to optimize parameters that provide an accurate representation of the experimental QD. We believe, the Green-Sellin-Zachor \cite{green69} is better suited to our present study as the parameter space is small (two-dimensional) which allows for an efficient search of the optimized parameters. \\

The electron-ion potentials obtained for each Rydberg series are further modified to account classically for the electron-electron interaction in the final state. Calculations performed within the DWBA framework for the ionization of argon in the equal energy sharing geometry
${\bf k}_a=-{\bf k}_b$, which is reasonably well documented both theoretically and experimentally, clearly validate our approach and shows significant improvements over previous treatments.

This paper is organized as follows: Section 2 is devoted to the presentation of the CFM, section 3 describes the optimization method we use based on the CFM. In section 4 we describe the results whereas section 5 contains our conclusion and discussion of the results.

\section{THE CANONICAL FUNCTION METHOD}

The canonical function method (CFM) \cite{kobeissi82} is a powerful means for solving the Radial Schr\"{o}dinger 
Equation (RSE). The mathematical difficulty of the RSE lies in the fact it is a singular boundary value 
problem. The CFM turns it into a regular initial value problem and allows the full determination of the 
spectrum of the Schr\"{o}dinger operator bypassing the evaluation of the eigenfunctions.\\

The Canonical Function Method (CFM) developed by Kobeissi \cite{kobeissi91} and his coworkers to integrate the RSE, consists in writing the general  solution as a function of the radial distance r in terms of two basis functions $\alpha(E;r)$ and $\beta(E;r)$ for some energy $E$. \\
 
Generally, one avoids using the wavefunction but if one needs
it, the functions $\alpha(E;r)$ and $\beta(E;r)$ are used to
determine the wavefunction at any energy with the expression:
\begin{equation}
\Psi(E;r)=\Psi(E;r_0) \alpha(E;r) + \Psi'(E;r_0) \beta(E;r)
\end{equation}
where $\Psi(E;r_0)$ and $\Psi'(E;r_0)$ are the wavefunction and its derivative at
some initial distance $r_0$. \\

At the selected distance $r_0$ a well defined set of initial conditions are satisfied by the canonical functions and their derivatives ie: $\alpha(E;r_0)=1$ with $\alpha'(E;r_0)=0$ and $\beta(E;r_0)=0$ with $\beta'(E;r_0)=1$.\\

The method of solving the RSE is to proceed from $r_0$ simultaneously towards the origin ($r=0$) and towards $\infty$. 
During the integration, the ratio of the canonical functions is monitored until
 saturation signaling the  stability of the eigenvalue spectrum.\\
 
The saturation of the $\frac{\alpha(E;r)}{\beta(E;r)}$ ratio with r yields a position independent eigenvalue function $F(E)$. The latter is defined with the help of two associated energy functions:
\begin{equation}
l_{+}(E)=lim_{r \rightarrow +\infty} -\frac{\alpha(E;r)}{\beta(E;r)}
\end{equation}

and:
\begin{equation}
l_{-}(E)= lim_{r \rightarrow 0} -\frac{\alpha(E;r)}{\beta(E;r)}
\end{equation}

The eigenvalue function is defined in terms of:

\begin{equation}
F(E)=l_{+}(E)-l_{-}(E)
\end{equation}

Its zeroes yield the spectrum of the RSE. An example of behaviour of $F(E)$ is
displayed in Figure 1. The eigenfunctions may be obtained for any $E=E_k$ where $E_k$ is a zero of $F(E)$.\\

\begin{figure}[htbp]
\begin{center}
\scalebox{0.6}{\includegraphics*{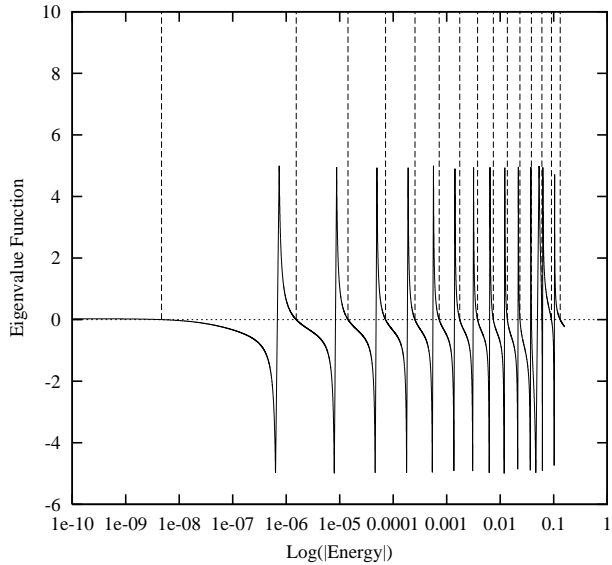}}
\end{center}
\caption{Typical behavior of the eigenvalue function with energy. The vertical lines indicate the eigenvalue position.}
\label{fig1}
\end{figure}

The speed and accuracy of the method have been tested and compared to standard integration algorithms 
in a variety of cases and for a wide of range of potentials. In addition the method has been generalised to 
the phase shift estimation \cite{kobeissi90} as well to the continuum coupled channel problem \cite{kobeissi91}.\\

In potential optimisation problems the method can only be of great value because of its speed, accuracy 
and ease of programming. Accordingly, one has to pick an appropriate optimisation method and use the 
CFM as a subroutine for a given parametric potential.

\section{OPTIMISATION PROCEDURE}

An extensive review of the applications of model potentials has been given by Hibbert  \cite{hibbert82}
and Aymar {\sl et al.} \cite{aymar96}. The functional form suggested by Green {\sl et al.} \cite{green69} is given by:
\begin{eqnarray}
&&V(r)= -(2/r) [(Z-1) \omega(r)+1],\nonumber\\
\mbox{with}\hspace{0.5cm}&& \omega(r)=1/[\epsilon_1{{\rm (exp}(r/\epsilon_2)-1})+1]
\label{eq5}
\end{eqnarray}
where $\epsilon_1$ and $\epsilon_2$ are parameters that are determined by the optimisation procedure.\\

The optimisation problem, at hand, is over-determined since the experimental set of energy levels might consist of tens of values whereas the potential depends only on two
numbers, namely $\epsilon_1$ and $\epsilon_2$. This over-determination allows us to use several criteria for the optimisation procedure and later on select the best one
that achieves results closest to experiment.\\

The optimization procedure consists of defining an objective function
and finding its minimum in the two-dimensional parameter space {$\epsilon_1$, $\epsilon_2$}.
The objective function is a quadratic consisting of the difference 
between some picked levels and those produced by the parametric potential through the
solution of the RSE with the CFM.\\

We adopted several strategies based on the following observations:
The quantum defect value is not stable for the low levels but tends
to reach a stable value when the energy increases. 
Despite the stability of the quantum defect for the higher levels, 
the experimental (and therefore) numerical accuracy decreases when
the energy increases.\\

Therefore a compromise should be achieved by selecting the levels in order to
define the objective function to minimize. \\

We found that a reasonable compromise should be based on the following operations that
differ with the selected rare gas: 

\begin{enumerate}
\item Pick some level and take the average of a number of higher ones.
\item Pick two high levels for which the Quantum defect has stabilised
within a given accuracy.
\item Pick a low level and a high one for which the Quantum defect has
already stabilised.
\end{enumerate}

All the above operations should yield roughly the same values for
the parameters before running the final check in order to test the
accuracy of the obtained eigenvalues.\\

The optimization program itself is based on a globally convergent
method for solving non-linear system of equations: the multidimensional
secant method developed by Broyden \cite{broyden}.
It is based on a fast and accurate method for the iterative evaluation
of the Jacobian of the objective function needed during the minimisation 
procedure.\\

It is a Quasi-Newton method that consists of approximating the Jacobian
and updating it with an iterative procedure. It converges superlinearly
to the solution like all secant methods.\\

There are several ways to perform the integration of the RSE on the basis of the CFM. 
One may use a fixed step scheme such as the explicit Runge-Kutta fourth order method (RK4) 
or a Variable Step with Control of Accuracy (VSCA). Optimization wise, the RK4 method is faster
 than VSCA but less accurate. To compromise, we first perform an initial search of the parameters 
with RK4 and then finalize the results using the VSCA method. Table \ref{tab1} shows the parameters
 obtained for the first three Rydberg series of argon, optimized using the experimental energy
 levels tabulated by Moore \cite{moore}. \\

\begin{table}[htbp]
\begin{center}
\begin{tabular}{|l|c|r|}
\hline
 Series & $\epsilon_1$ &  $\epsilon_2$\\
\hline   
   l=0 &    3.625 & 1.036\\
   l=1  &   3.62  & 1.06\\
   l=2 & 3.6344  & 1.036 \\
\hline     
\end{tabular}
\end{center}
\caption{Czylik-Green parameters for the first three Rydberg series of Argon..}
\label{tab1}
\end{table}

\begin{table}[htbp]
\begin{center}
\begin{tabular}{|l|c|r|}
\hline
Experimental levels & Calculated levels (RK4) & Calculated levels (VSCA) \\
\hline
   -0.309522&  -0.214082&  -0.310563\\
   -0.124309&   -9.87422E-02&  -0.124506\\
    -6.76780E-02&   -5.69743E-02&   -6.76904E-02\\
    -4.25540E-02&   -3.70845E-02&   -4.25546E-02\\
    -2.92210E-02&   -2.60593E-02&   -2.92238E-02\\
    -2.13080E-02&   -1.93126E-02&   -2.13062E-02\\
    -1.62200E-02&   -1.48846E-02&   -1.62210E-02\\
    -1.27620E-02&   -1.18223E-02&   -1.27614E-02\\
    -1.03020E-02&   -9.61654E-03&   -1.03015E-02\\
    -8.49000E-03&   -7.97521E-03&   -8.49011E-03\\
    -7.11800E-03&   -6.72092E-03&   -7.11771E-03\\
\hline     
\end{tabular}
\end{center}
\caption{Comparison between the experimental and calculated energy levels of the Rydberg series of Argon. 
The levels calculated with the CFM are obtained either with fixed step (RK4) or variable step (VSCA) integration. All values in Rydbergs.}
\label{tab2}
\end{table}

\begin{table}[htbp]
\begin{center}
\begin{tabular}{|l|c|r|}
\hline
Experimental QD & Calculated QD (RK4) & Calculated QD (VSCA) \\
\hline
0.202561 & 0.838726 & 0.205576 \\
0.163723 & 0.817645 & 0.165967 \\
0.156063 & 0.810516 & 0.156415 \\
0.152366 & 0.807174 & 0.152400 \\
0.150046 & 0.805324 & 0.150326 \\
0.149399 & 0.804191 & 0.149110 \\
0.148104 & 0.803444 & 0.148345 \\
0.148016 & 0.802940 & 0.147808 \\
0.147664 & 0.802574 & 0.147425 \\
0.147091 & 0.802298 & 0.147161 \\
0.147199 & 0.802084 & 0.146957 \\
\hline     
\end{tabular}
\end{center}
\caption{ Comparison between the experimental and calculated QD of the Rydberg series of Argon. 
The levels calculated with the CFM are obtained either with fixed step (RK4) or variable step (VSCA) integration.}
\label{tab3}
\end{table}

To judge the accuracy of our optimization we compare in tables \ref{tab2} and \ref{tab3} our results for the energy levels and corresponding quantum
 defect obtained 
using the RK4 and VSCA integration scheme with those of Czydlik {\sl et al.} \cite{szydlik74}. The results displayed in 
Tables \ref{tab2} and \ref{tab3} clearly favor, as expected, the VSCA integration scheme. In contrast to the VSCA, the RK4 
integration scheme is limited to fourth order accuracy. Table \ref{tab3} further shows the sensitivity of the QD 
to the numerical values of the calculated energy levels. Despite small differences between the energy values 
obtained with the RK4 and VSCA methods (Table \ref{tab2}), the corresponding QD's largely differ. This is the 
main motivation for using QD rather than energy levels in our optimization procedure.

\section{Scattering method and results}

We now turn to the implementation of the optimized pseudo-potentials to the case of (e,2e) ionization in the equal
 energy sharing case (${\bf k}_{a} =- {\bf k}_{b}$). Within a standard DWBA model and when a symmetric kinematic 
is considered (i.e. equal energy sharing between the two outgoing electrons) the exact unsymmetrized $T$-matrix
 element is approximated as \cite{mccarthy, weigold}:

\begin{equation}
\langle {\bf k}_{a} {\bf k}_{b} \Phi^{ion}_{J_{i},L_{i},M_{i}}|T|\Phi^{atom}_{J,L,M}{\bf k}_{0} \rangle \equiv
\langle \chi^{-}({\bf k}_{a})\chi^{-}({\bf k}_{b})|V|\phi_{L,M} \chi^{+}({\bf k}_{0}) \rangle ,
\end{equation}

and both final state distorted waves, $\chi^{-}({\bf k}_{a})$ and  $\chi^{-}({\bf k}_{b})$ are chosen as eigenfunctions of the electron-ion system. In this situation, the exact collision state with final state boundary condition, $ \Psi^{(-)}({\bf k}_{a},{\bf k}_{b}) $, 
is approximated as the product of two outgoing distorted waves. As for the calculation of the incoming distorted wave, the Furness-McCarthy local exchange approximation which, valid for scattering energies greater than 25eV \cite{riley}, 
is commonly used is the calculation of the outgoing distorted waves. Furthermore, the distorting potential
 is purely Coulombic asymptotically, and thereby neglects any e-e correlation in the final state.

It is well known \cite{pan,peterkop} that the product of two Coulomb waves satisfy the proper asymptotic condition 
provided the use of effective charges that satisfy the condition
\begin{equation}
\frac{Z_a^{eff}}{k_a}+\frac{Z_b^{eff}}{k_b}=\frac{1}{k_a}+\frac{1}{k_b}-\frac{1}{|{\bf k}_a-{\bf k}_b|}.
\label{pet1}  
\end{equation}
In the present case, we focus on the equal energy sharing geometry ${\bf k}_a=-{\bf k}_b$ and impose on the effective 
charge the condition
\begin{equation}
Z_a^{eff}=Z_b^{eff}
\label{cond1}
\end{equation}
Accordingly, we modify the form of our pseudo-potentials and replace in eq. (5), and for all $r$ values, the unit charge by
an effective charge $Z^{eff}$ which satisfy (\ref{pet1}) and (\ref{cond1}):
\begin{equation}
Z^{eff}=\frac{3}{4}.
\end{equation} 
This choice of screening charge insures that each outgoing electron experiences the exact classical force asymptotically. 
It was further suggested that the use of such effective charges suffices to take into account the electron correlation for 
the ionization of rare gas atoms in equal energy sharing geometry \cite{pan}. \\

In the spirit of a DWBA method based on pseudo potentials and to describe the ionization process
 of argon at energy beyond the validity of the Furness-McCarthy local exchange potential (below 25eV),
 we use the parametric form suggested by Jungen {\sl et al.} \cite{roche} to describe the incoming distorted wave. 
This pseudo-potential is adjusted to reproduce the {\sl ab-initio} elastic phase shift of 
McEachran {\sl et al.} \cite{maceachran} up to an energy of 30eV and includes static, exchange and 
polarization effects. At energy above 30eV, the Furness-McCarthy local exchange potential 
is used in the entrance channel.

\begin{figure}[htbp]
\begin{center}
\scalebox{0.8}{\includegraphics*{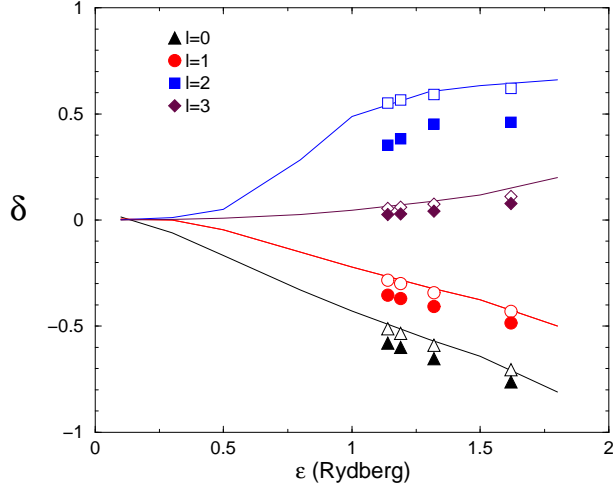}}
\end{center}
\caption{Comparison of the {\sl ab-initio} elastic scattering phaseshift of McEachran {\sl et al.} (solid line) with the phaseshift obtained using the parametric potential of Jungen {\sl et al.} (opaque symbols) and the Furness-McCarthy local exchange potential(filled symbols).}
\label{fig2}
\end{figure}

To gauge the improvement introduced using such a parametric potential to describe the incoming projectile, we show in Fig 2 a comparison of the elastic scattering phaseshift obtained using the pseudo-potential of Jungen {\sl et al.}, using the Furness-McCarthy local exchange approximation and the {\sl ab-initio} results of McEachran {\sl et al.} which leads to scattering parameters in good agreement with the experimental values. The significant improvement introduced in the calculation of the elastic scattering phaseshift, and consequently in the determination of the radial part of the incoming distorted wave, strongly suggests that a superior description of the ionization process is to be expected.

\begin{figure}[htbp]
\begin{center}
\scalebox{0.9}{\includegraphics*{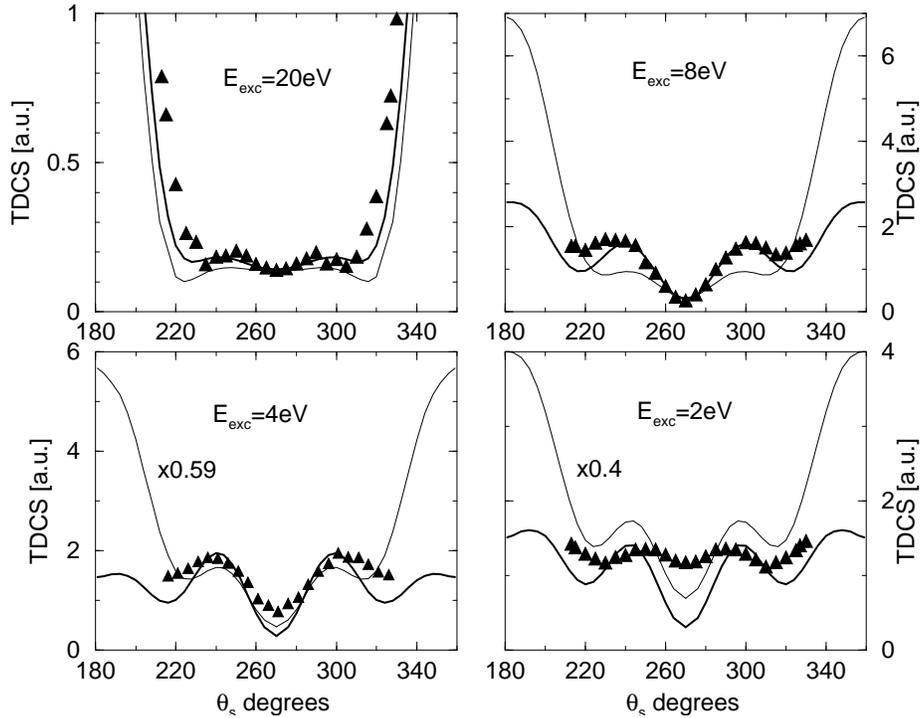}}
\end{center}
\caption{ Behavior of the TDCS as a function of scattering angle in equal energy sharing geometry for various values of the excess energy as indicated on each graph. DWBA calculations (thin line), DWBA calculations + pseudo-potentials (thick line), experiment (triangle).}
\label{fig3}
\end{figure}

Fig 3 shows the variation of triply differential cross section (TDCS) as a function of the scattering angle for various values of excess energy, equally shared between the two outgoing electrons. The results obtained using the pseudo-potentials defined in the previous sections for the incoming and outgoing distorted waves are compared to calculations where the Furness McCarthy local-exchange potential \cite{mccarthy} is used in both the entrance and the exit channels and to the experimental measurements available \cite{rouvellou}. Since the experimental measurements are not absolute, they have been systematically normalized to the DWBA calculations at a scattering angle $\theta$ = 270 degrees except for an incident energy of $E_{i}$ = 17.8eV (i.e. $E_{exc}$ = 2eV) and $E_{i}$ = 19.8eV (i.e. $E_{exc}$ = 4eV) where the experiment is normalized to obtain the best visual fit to the theory.

For the highest incident energy considered here, $E_{0}$ = 35.8 eV (i.e. $E_{exc}$ = 20eV ), little differences in the magnitude of the TDCS is noticeable between the two theoretical models. In contrast, the use of the pseudo-potentials calculated in the previous section clearly improves the prediction of the shape of the TDCS. This is particularly the case of the angular region between 230 and 320 degrees. Although improved, the TDCS is however still slightly underestimated outside this particular angular range. We stress that at this particular energy, the Furness-McCarthy local exchange approximation is used in the calculation of the incoming distorted wave. Further, this result also validates our procedure for calculating the outgoing distorted wave for an energy interval (i.e 35.8eV for the incoming electron) where the Furness-McCarthy local exchange potential is still in its lower energy limit of validity. Similar remarks pertain for the magnitude of the TDCS at the next lower incident energy ($E_{exc}$ = 8eV ). At this particular regime where the incoming distorted wave is now also described using pseudo-potentials, our optimized potentials largely improves the description of the experimental TDCS, especially in the angular region between 240 and 300 degrees.

At the two lowest energies considered here, marked differences are now noticeable between the TDCS obtained using the two different theoretical models (i.e DWBA and DWBA+pseudo-potentials). First, the overall magnitude of the TDCS is largely reduced when parametric potentials are used. Additional calculations (not shown here) indicate that the use of effective charges is largely at the origin of this reduction. Unfortunately, the improvement introduced can not be gauged against the experimental data as the measurements are not absolute. The reduction in amplitude of the variation of the TDCS where experimental measurements are available can be traced back to the influence of the pseudo-potential combined with the use of the effective charges in the exit channel \cite{mazevet}. Finally, when the pseudo-potentials of Jungen {\sl et al.} are introduced to describe the entrance channel distorted wave little improvements are observed in the calculations of the TDCS where experimental measurements are available. Its effects on the calculated cross sections are mainly felt at the lowest and highest angular region where experimental measurements are not available.
  
\section{CONCLUSIONS}

In this work, we showed that QD information provides means to select and tune a class of pseudo-potentials. These can subsequently be used in ionization studies by electron impact with proper modifications to account for the electron correlation in the final state. Ionization of argon, in the equal energy sharing geometry, shows that this procedure provides an efficient treatment of the final state interactions in the energy region where the commonly used Furness-McCarthy local exchange potential is no longer valid, and yet allows for a much simpler alternative to the complete Hartree-Fock treatment. This might turn out to be particularly valuable when considering the ionization of molecular systems involving low energy electrons. We finally stress that the use of QD information, and consequently the application of the present method, would not be meaningful for energy above 30eV.

\section{ACKNOWLEDGMENTS} 

Helpful correspondance, regarding mathematical aspects of the RSE, with Dr. Jeff 
Cash are gratefully acknowledged.


\begin{thebibliography}{99}
\bibitem{jungen} C. Jungen, Molecular Applications of Quantum Defect Theory,
Institute of Physics Publishing (1996).
\bibitem{aymar96} M.~Aymar, C.~H. Green, and E.~Luc-Koenig, Rev. Mod. Phys. {\bf 68}, 1015 (1996).
\bibitem{furness} J. B. Furness and I. E. McCarthy J. Phys. B 6 2280 (1973).  
\bibitem{riley} M.E. Riley and D.G. Trular,J. Chem. Phys. 63, 2182 (1975).
\bibitem{winkler} K. D. Winkler, D. H. Madison and H. P. Saha J. Phys. B 32, 1987 (1999).
\bibitem{green69} A.~E.~S. Green, D.~L. Sellin and  A.~S.Zachor, Phys. Rev. {\bf 184}, 1 (1969).
\bibitem{kobeissi82} H.~Kobeissi, J. Phys. B {\bf 15}, 693 (1982).
\bibitem{tannous99} C.~Tannous, K.~Fakhreddine and J.~Langlois, J. Phys. IV France {\bf 9} Pr6-71 (1999).
\bibitem{kobeissi90} H.~Kobeissi, K.~Fakhreddine and M.~Kobeissi, Int. J. Quantum Chemistry, Vol XL, 11 (1990).
\bibitem{kobeissi91} H.~Kobeissi, K.~Fakhreddine  J. Physique II (France) {\bf 1}, 38 (1991).
\bibitem{hibbert82} H.~Hibbert, Adv. At. Mol. Phys. {\bf 18} 309 (1982).
\bibitem{szydlik74} P.~P. Szydlik and A.~E.~S.Green, Phys. Rev. {\bf A9}, 1885 (1974).
\bibitem{klapisch71}  M.~Klapisch, Comp. Phys. Comm. {\bf 2},  239 (1971).
\bibitem{broyden} Broyden, CG: Mathematics of Computation, 19, 557 (1965). 
\bibitem{moore} C. E. Moore, Atomic energy levels, NBS Publications (1971). 
\bibitem{mccarthy} I. E. McCarthy, Aust. J. Phys. 48, 1 (1995). 
\bibitem{weigold} I. E. McCarthy and E. Weigold, Electron-atom collisions, Cambridge University Press (1995). 
\bibitem{pan} C. Pan and A. F. Starace, Phys Rev A 45, 4588 (1992). 
\bibitem{peterkop} R. K. Peterkop, Theory of Ionization of Atoms By Electron Impact
 (Colorado Associated University Press, Boulder) (1977). 
\bibitem{roche} C. Jungen, A.L. Roche, M Arif, Phil. Trans. A, 355, 2520 (1997).
\bibitem{maceachran} R. P. McEachran, A. D. Stauffer, J. Phys. B 16, 4023 (1983). 
\bibitem{rouvellou} B. Rouvellou, S. Rioual, J. Roeder, A. Pochat, J. Rasch, C. T. Whelan, H. R. J. Walters 
and R. J. Allan, Phys. Rev. A 57, 3621 (1998).
\bibitem {mazevet} S. Mazevet, K. Fakhreddine, G. Nguyen Vien, R. J. Tweed, and J. Langlois, in preparation.
\end{thebibliography}
\end{document}